\documentclass[aps,pra,multicolumn,twocolumn]{revtex4}
\usepackage{graphicx}
\usepackage{mathrsfs}
\usepackage{amssymb}
\usepackage{amsmath}
\usepackage{extarrows}

\begin{document}

\title{Entangled States of Two Quantum Dots Mediated by  Majorana Fermions}

\author{Z. C. Shi$^{1,2}$, W. Wang$^{1}$,  and X. X. Yi$^{1}$\footnote{Corresponding  address:
yixx@nenu.edu.cn}} \affiliation{$^1$ Center for Quantum Sciences and
School of Physics, Northeast
Normal University, Changchun 130024, China\\
$^2$ School of Physics and Optoelectronic Technology\\
Dalian University of Technology, Dalian 116024, China}

\begin{abstract}
With the assistance of a pair of Majorana fermions, we propose
schemes to entangle two quantum dots by Lyapunov control in the
charge and spin degrees of freedom. Four different schemes are
considered, i.e., the teleportation scheme, the crossed Andreev
reflection scheme, the intra-dot spin flip scheme, and the scheme
beyond intra-dot spin flip. We demonstrate that the entanglement can
be generated by modulating the chemical potential of quantum dots with
square pulses, which is easily  realized in practice. In contrast to
Lyapunov control, the preparation of entangled states by adiabatic
passage is also discussed. There are two  advantages in the scheme
by Lyapunov control, i.e., it is flexible to choose a control
Hamiltonian,  and the control time is much short with respect to the
scheme by adiabatic passage. Furthermore, we find that the results
are quite different by different  adiabatic passages in the scheme
beyond intra-dot spin flip, which can be understood as an effect
of quantum destructive interference.
\end{abstract}

 \maketitle

\section{Introduction}

Majorana fermions (MFs), which  are predicted to exist at the
boundary or in the vortex core of a topological superconductor (TS),
have been widely studied recently  both in theories
\cite{kitaev01,qi11,alicea12,beenakker13} and experiments
\cite{das2012,mourik12,rokhinoson12,lee13,perge14}. For example, MFs
have been applied into topological quantum computation
\cite{kitaev03,nayak08} due to their robustness against
perturbations.

From the other aspect, two spatially separated MFs can form a Dirac
fermion. This non-locality feature can offer the MFs an
opportunity as the medium to entangle quantum systems such as quantum
dots (QDs).
The advantages to entangle QDs via MFs are twofold. Firstly, conventional schemes to
entangle QDs are limited by spatial distance due to direct proximity coupling with each other \cite{loss98,petta05,koppens05}. As a result it is difficult to manipulate quantum states in such a small distance. The entanglement preparation
mediated by MFs can effectively solve this drawback. Secondly, from the side of
topological quantum computation, the MFs are difficult to couple
together as the wave-function overlap of two MFs decays
exponentially with spatial distance between them.
Additionally, braiding operations solely are insufficient in the universal quantum computation for Majorana-based qubits \cite{nayak08}.
Those difficulties can be availably overcomed in hybrid systems \cite{hassler11,sau10a,jiang11,bonerson11,leijnse12a,kovalev14,xue14}, where the combination of the robustness of topological qubits (e.g., Majorana-based qubit) and the universality of conventional qubits (e.g., QD-based qubit, flux qubit, etc.) can form an universal computation.

The adiabatic passage, which is widely applied to quantum
information processing, is one way to prepare entanglement.
The main idea is as follows. Given a quantum system that its ground
state is separable  and easy to prepare, one can adiabatically
manipulate some physical parameters such that the system evolves
into the ground state of the new Hamiltonian, which is the target
entangled state. Since the adiabatic dynamics is protected, it is
then immune to some types of perturbations. However, the price one
should  pay is the long time needed to finish the evolution. To
overcome this difficulty, a shortcut to adiabaticity
\cite{demirplak03,berry09} is introduced. The key point is that the
time-dependent Hamiltonian $H_1(t)$ is brought into the
Schr\"odinger equation. Nevertheless it is always hard to implement
such a Hamiltonian for most systems, and $H_1(t)$ may not exist in
complicated systems. This gives rise to a question---are there any
other control strategies for the quantum system better than
adiabatic passage?

In this work, we will focus on this issue. We propose to entangle
two QDs by Lyapunov control, which has been   employed in
manipulating quantum states
\cite{beauchard07,coron09,wang09,wang10a,yi09,shi15}. The basic
principle of Lyapunov control is to design control fields to drive a
quantum system to approach the target state. Note that the target
state must be a steady state such that the system cannot evolve when
control fields vanish. In order to design control fields, a Lyapunov
function has to be defined. To be specific, one first defines   a
Lyapunov function $V$ for the given target state $|\psi_T\rangle$.
Then by restricting non-positivity of the time derivative of the
Lyapunov function $V$ (i.e., $\dot{V}\leq0$), one can obtain the
control fields $f_k(t).$ Driven by these control fields, the system
would evolve to the target state $|\psi_T\rangle$ with time.

The system we considered is a  hybrid quantum system primarily consisting of
two QDs and a TS wire with MFs. The two QDs are well spatially
separated so that they do not have direct interaction, but they are
coupled to a common TS wire. The entanglement between the two QDs is
then induced via the non-locality property of the Dirac fermion
defined  by MFs. We adopt the Lyapunov control to explore
entanglement generation in the following. For comparison, the
entanglement generated by adiabatic passage is also presented and
discussed.

\section{Model}

\begin{figure}[h]
\centering
\includegraphics[scale=0.45]{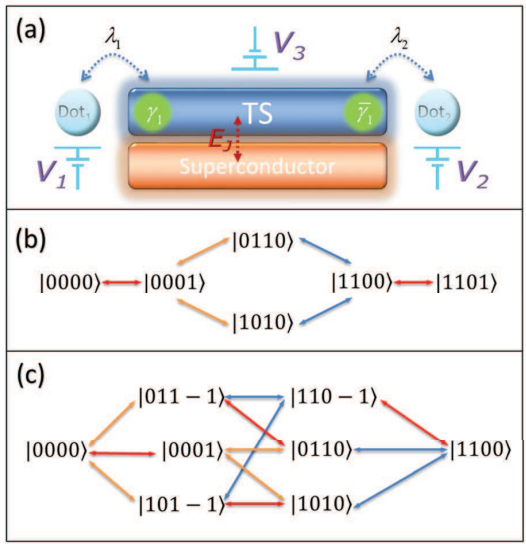}
\caption{(a) Schematic illustration of the setup.  A TS wire with
strong spin orbit interaction is  coupled to two QDs and a bulk
superconductor via tunneling. The TS wire is deposited on a s-wave
superconductor. When the TS wire is in the topological nontrivial
phase due to  the proximity effect and a Zeeman field, there exist a
pair of MFs at the ends of TS wire. The chemical potentials of QDs
can be changed  by  the gate voltage $V_{1,2}$,  while the gate
charge $n_g$ depends  on the gate voltage $V_3$ of the capacitor
(not shown in figure). (b) The transitions between distinct states with $n_g=1$.
The red line represents the Josephson coupling while the orange and
blue lines describe  the ``anomalous'' and ``normal'' tunneling
process, respectively. (c) The transition between distinct states with
$n_g=0$.} \label{fig:01}
\end{figure}

The system under consideration  is illustrated  in Fig.
\ref{fig:01}(a), which consists of a TS wire  coupled  to two QDs
and another bulk superconductor via tunneling.  The free
Hamiltonian of such a system contains two parts. The first part of
the free Hamiltonian is for the  TS wire ($\hbar=1$),
\begin{eqnarray}
H_{0}^{TS}=\mathcal{H}_{TS}+\mathcal{H}_{TS}',
\end{eqnarray}
where $\mathcal{H}_{TS}=\sum\xi_k\gamma_k^{\dag}\gamma_k$. The
energy  spectrum is $\xi_k$,  and the Bogoliubov quasiparticle
operators $\gamma_{k}$ are related to the electron operators $c(x)$
of the TS wire in the standard way,  where $x$ denotes the electron
coordinate in the TS wire \cite{alicea12}. Note that we have dropped
the spin subscript in this section as the spin degeneracy in the TS
wire is broken due to Zeeman effect, which suggests that we might
consider only one spin direction. Considering the situation that the
TS wire is in the topological nontrivial phase, there is a pair of
MFs $\gamma_1$ and $\bar{\gamma}_1$ at the ends of TS wire
\cite{oreg10,lutchyn10}. This two zero-energy MFs can form a
non-local  Dirac fermion, i.e.,
$f=\frac{\gamma_1+i\bar{\gamma}_1}{2}$, where $f$ is the
annihilation operator of Dirac fermion. When  the energy
scale of TS wire is smaller than the superconductor gap, there are
no other quasiparticle excitations except for the zero-energy MFs in
the TS wire. As a result we can safely ignore the Hamiltonian
$\mathcal{H}_{TS}$ in the model. Note that when the TS wire is of
mesoscopic size and is linked to a capacitor earth grounded, there
exists an additional Hamiltonian $\mathcal{H}_{TS}'$ describing  the
finite charging energy. The corresponding Hamiltonian takes
\cite{fu10}
\begin{eqnarray}  \label{2a}
\mathcal{H}_{TS}'=E_{c}(2\hat{N}_c-n_g+n_f)^2,
\end{eqnarray}
where ${N}_c$ is the number of Cooper pairs, $n_g$ denotes the
dimensionless gate charge determined by the gate voltage $V_3$, and
$n_f =f^{\dag}f$ stands for the number of Dirac fermion formed by
MFs. The single electron charging energy is $E_c=\frac{e^2}{2C}$
with capacitance $C$. It has been shown that the single electron
charging energy $E_c$ plays a key role in the long-range
entanglement generation of two QDs \cite{plugge15}.

The second part of the free Hamiltonian is for the  QDs,
\begin{eqnarray}   \label{3a}
H_0^{QD}=\sum_{n=1}^{2}\epsilon_{n}d_{n}^{\dag}d_{n},
\end{eqnarray}
where $d_n$ and $d_n^{\dag}$ are the annihilation and creation
operations of electrons in the $n$-th QD. $\epsilon_n$ denotes the
chemical potential which can be changed by the gate voltage $V_n$
($n=1,2$). We should emphasize that  we have assumed both QDs   in
the Coulomb block regime in Hamiltonian $H_0^{QD}$, i.e., the
electron can only occupy single fermion level. This requires that
$U_n$ [see in Eq. (\ref{9})] is   large compared to the other
relevant energy in the system.

Next we turn to the interaction Hamiltonian of system. The
first term  describes  the Cooper pair exchange between the TS wire
and the bulk superconductor,
\begin{eqnarray}
H_e^{TS}=E_{J}\cos{\hat{\phi}}=\frac{E_{J}}{2}(e^{-i\hat{\phi}}+e^{i\hat{\phi}}).
\end{eqnarray}
$E_J$ is the Josephson coupling and $\phi$ is the phase difference
between the two superconductors. Here the phase of bulk
superconductor have been  set to zero for simplicity. The operator
$e^{-i\hat{\phi}}$ ($e^{i\hat{\phi}}$) represents the creation
(annihilation) of a Cooper pair, i.e.,
$e^{-i\hat{\phi}}=\sum_{N_c}|N_c+1\rangle\langle N_c|$. The number
operator of Cooper pairs  and  the phase of superconductor are
canonically conjugate,  i.e., $[\hat{N}_c,
e^{-i\hat{\phi}}]=e^{-i\hat{\phi}}$.

The second  term of interaction Hamiltonian describes   the
electron tunneling between the TS wire and the QDs. For later use,
we  write down the electron operator $c(x)$ in terms of
quasiparticle operator $\gamma_k$ in the TS wire, i.e.,
$c(x)=g_1\gamma_1+\bar{g}_1\bar{\gamma}_1+\cdots$, where $g_k$ is
the wave function of spatial coordinate. As we have mentioned
before, there are no other quasiparticles except for the  MFs in the
TS wire. This allows us to consider the first two terms in the
expression of $c(x)$, i.e.,
$c(x)=g_1\gamma_1+\bar{g}_1\bar{\gamma}_1$. In addition, if the
length of TS wire is large enough, there are no overlap between
$g_1$ and $\bar{g}_1$, i.e., no coupling between $\gamma_1$
($\bar{\gamma}_1$) and $d_2$ ($d_1$). As a consequence, the
effective Hamiltonian describing tunnel coupling between  electron
in the QDs and  Dirac fermion formed by  the MFs is given by
substituting $c(x)$ into the bare tunneling terms $\sim\int
dxt_nd_n^{\dag}c(x)+h.c.$ \cite{zazunov11,hutzen12,didier13},
\begin{eqnarray}
H_t^{TQ}=\sum_{n=1}^{2}\lambda_{n}[f^{\dag}+(-1)^{n-1}e^{-i\hat{\phi}}f]d_{n}+h.c.,
\end{eqnarray}
where $\lambda_n$ represents the coupling strength. In the
derivation  of  Hamiltonian $H_t^{TQ}$, we have considered  charge
conservation since it cannot create or annihilate  charge $2e$
without any energy cost in the TS wire with $E_c\neq0$. In later
discussion, the terms $f^{\dag}d_n$ (or $d_n^{\dag}f$) and $fd_n$
(or $d_n^{\dag}f^{\dag}$) are referred to the ``normal'' and
``anomalous'' tunneling process, respectively.

As we study the long-range entanglement generation in this
paper, the length of TS wire is considered  so sufficiently long that
there does not exist direct coupling between the two MFs. Therefore the
interaction Hamiltonian $H_i^{MF}=i\lambda_f\gamma_1\bar{\gamma}_1$
has also been neglected throughout this work.

Collecting  these terms, we can write down the Hamiltonian for the
whole system
\begin{eqnarray}
H_0=H_0^{TS}+H_0^{QD}+H_e^{TS}+H_t^{TQ}.
\end{eqnarray}
In order to prepare maximally entangled state of the two QDs, e.g.,
Bell states, we choose $\epsilon_1=\epsilon_2=\epsilon$ and
$\lambda_1=\lambda_2=\lambda$, and work in the large-$E_c$ limit
(compared  to the parameters $\epsilon$ and $\lambda$) which
can always be satisfied by changing the size of TS wire. Since
the total fermion parity, which is quantized  by the number of the
electrons in two QDs plus the Dirac fermion formed by MFs, is
conserved in this model, we will study the even-parity case in the
following. The extension from the even-parity case to the
odd-parity case is straightforward.

\section{Teleportation scheme}

The teleportation (TP) refers to nonlocal transfer of fermions
across the TS wire \cite{plugge15,fu10}. As an example, we briefly present
the physical process that electrons transfer from the
first QD to the second QD via the TS wire. When an electron
tunnels from the first QD to the TS wire, the charges in the TS
wire increase one unit with energy cost $E_c$. If $E_c$  can not match
$\epsilon$, the tunneling process is  virtual. In order to keep the
conservation of charge in the TS wire, the electron may transfer
to the second QD. We first consider a case in which the Josephson
coupling $E_J=0$ and the gate charge $n_g=1$. The ground
state corresponds to that the number of Cooper pair $N_c=0$, Dirac
fermion $n_f=1$, and the electrons $n_1+n_2=1$ in the even-parity case
[cf., Eqs.(\ref{2a})-(\ref{3a})]. Namely, the ground states
are $|0110\rangle$ and $|1010\rangle$. We have employed the notation
$|n_1n_2n_f N_c\rangle$ to describe the system state, where $n_1$
($n_2$) is the number of electrons in the first (second) QD, $n_f$
is the number of Dirac fermion formed by MFs, and $ N_c$ is the
number of Cooper pairs in the TS wire. $|0001\rangle$ and
$|1100\rangle$ are the low  excited states coupled directly  to the
ground state. We ignore the higher-energy excited states due to
large gaps to the ground state. Since the entanglement generation is
based on the nonlocality of the Dirac fermion defined  by MFs [cf.,
Fig. \ref{fig:01}(b)], we refer this proposal  to teleportation
scheme.

According to the transitions  shown in Fig. \ref{fig:01}(b), the
Hilbert space is spanned by \{$|0001\rangle$, $|0110\rangle$,
$|1010\rangle$, $|1100\rangle$\}. In this space, the Hamiltonian
$H_0$ can be written as a $4\times4$ matrix, i.e.,
\begin{eqnarray}
H_0=\left(
  \begin{array}{cccc}
    E_c & -\lambda & \lambda & 0 \\
    -\lambda & \epsilon & 0 & \lambda \\
    \lambda & 0 & \epsilon & \lambda \\
    0 & \lambda & \lambda & E_c+2\epsilon \\
  \end{array}
\right).
\end{eqnarray}
The system eigenstates then can be found analytically,
\begin{eqnarray}  \label{2}
|E_1\rangle&=&\mathcal{N}_1(|1010\rangle-|0110\rangle-A_1|0001\rangle),          \nonumber\\
|E_2\rangle&=&\mathcal{N}_2(|1010\rangle-|0110\rangle+A_2|0001\rangle),          \nonumber\\
|E_3\rangle&=&\mathcal{N}_3(-\frac{A_3}{2}|1010\rangle-\frac{A_3}{2}|0110\rangle+|1100\rangle), \nonumber\\
|E_4\rangle&=&\mathcal{N}_4(\frac{1}{A_3}|1010\rangle+\frac{1}
{A_3}|0110\rangle+|1100\rangle),
\end{eqnarray}
and the corresponding eigenvalues are given by $E_1=\epsilon-\lambda
A_1$, $E_2=E_c+\lambda A_1$, $E_3=E_c+2\epsilon-\lambda A_3$,
$E_4=\epsilon+\lambda A_3$, where
$A_1=\frac{\epsilon-E_c+\sqrt{(E_c-\epsilon)^2+8\lambda^2}}{2\lambda}$,
$A_2=\frac{E_c-\epsilon+\sqrt{(E_c-\epsilon)^2+8\lambda^2}}{2\lambda}$,
$A_3=\frac{E_c+\epsilon+\sqrt{(E_c+\epsilon)^2+8\lambda^2}}{2\lambda}$.
$\mathcal{N}_j$ ($j=1,2,3,4$) is the normalization constant. One can
find in Eq.(\ref{2}) that $|E_1\rangle\simeq|0001\rangle$ if
$\epsilon\gg E_c$, while if $\epsilon\ll E_c$,
$|E_1\rangle\simeq\frac{1}{\sqrt{2}}(|10\rangle-|01\rangle)|10\rangle=|\psi_T\rangle,$
which is the Bell state of the two QDs. This observation paves the
way towards preparing the Bell state $|\psi_T\rangle$ by adiabatic
passage. To be explicit, we first initialize  the system in state
$|0001\rangle$. Then we adiabatically decrease the chemical
potential $\epsilon$. The system would stay in the eigenstate
$|E_1\rangle$, and arrive at the Bell state $|\psi_T\rangle$
finally. Similarly, when the initial state is $|1100\rangle$, the
system would follow the eigenstate $|E_3\rangle$ to arrive at the
Bell state.

\begin{figure}[h]
\centering
\includegraphics[scale=0.45]{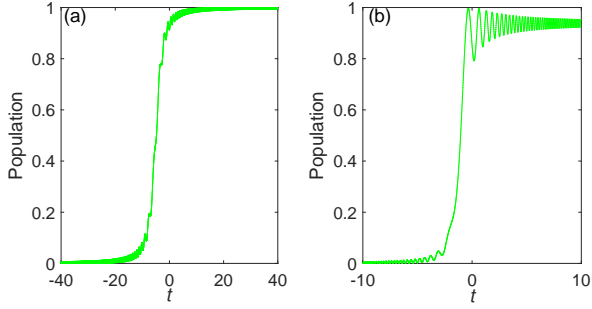}
\caption{Population of the Bell state $|\psi_T\rangle$ as a function
of time, where the population is defined as $|\langle
\psi(t)|\psi_T\rangle|^2$. We have set the chemical potentials of
the two QDs decrease with time linearly, e.g.,
$f(t)=-\frac{40}{T}t+20$. As a result the total manipulation time is
$2T$. All parameters are in units of tunnel coupling $\lambda$.
$E_c=30$, $\epsilon=5$. (a) $T=40$. (b) $T=10$.}  \label{fig:02}
\end{figure}

In order to compare adiabatic passage with Lyapunov control, we
reformulate the aforementioned results from the point  of quantum
control. That is, consider  the chemical potentials of two QDs being
manipulated in adiabatic passage, the control Hamiltonians read,
\begin{eqnarray}  \label{3}
H_{1}&=&d_{1}^{\dag}d_{1},  \nonumber\\
H_{2}&=&d_{2}^{\dag}d_{2}.
\end{eqnarray}
Hence the system evolution is governed by the following
Schr\"odinger equation,
\begin{eqnarray}
i|\dot{\psi}\rangle=[H_0+\sum_{k=1}^{2}f_{k}(t)H_k]|\psi\rangle,
\end{eqnarray}
where $f_{k}(t)$ denotes the chemical potential changing with time.
We refer $f_{k}(t)$  as control field hereafter, and set two control
fields the same to simplify  experimental realizations, i.e.,
$f_1(t)=f_2(t)=f(t)$. We plot in Fig. \ref{fig:02} the time
evolution when linearly decreasing the chemical potentials of two
QDs . It suggests that one can indeed achieve the Bell state
$|\psi_T\rangle$ when the control fields $f(t)$ change sufficiently
slowly with time [for instance, in Fig. \ref{fig:02}(a)], but it
takes long manipulation time. With increasing the change rate of
chemical potentials,  shown in Fig. \ref{fig:02}(b), the performance
gets worse due to the breakdown of adiabatic condition. Therefore
the fact that it is difficult to decrease the manipulation time
becomes a bottleneck for adiabatic passage.

\begin{figure}[h]
\centering
\includegraphics[scale=0.5]{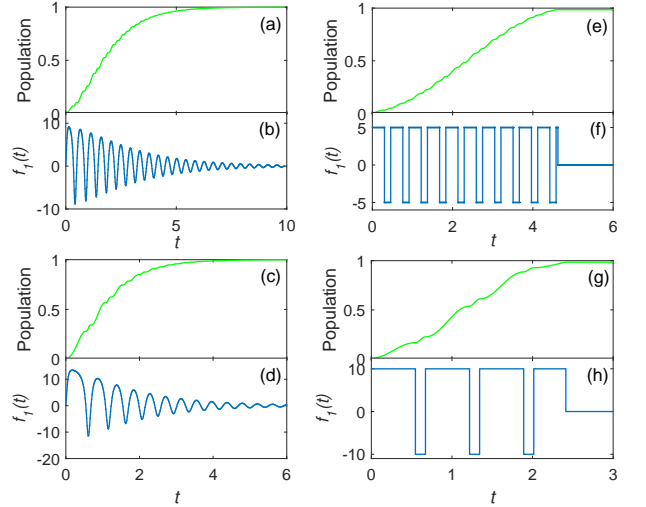}
\caption{The system evolution as a function of time. Panel (a) is
the population of Bell state $|\psi_T\rangle$ and the corresponding
control field $f_1(t)$ is depicted in panel (b). The meaning of
panels (c)-(d), panels (e)-(f), and panels (g)-(h) are the same as
panels (a)-(b)  except for different $B_1$ or $F$. All parameters
are in units of the tunnel coupling $\lambda$. $E_c=20$,
$\epsilon=5$. (b) $B_1=200$. (d) $B_1=300$. (f) $F=5$. (h) $F=10$. }
\label{fig:03}
\end{figure}

Next we employ Lyapunov control to speedup the entanglement
generation.
In order to determine the form of control fields, one has to
choose a Lyapunov function $V$ first, e.g.,
\begin{eqnarray}\label{6}
V=1-|\langle \psi_T|\psi\rangle|^2,
\end{eqnarray}
where $|\psi_T\rangle$ is the target state (i.e., the Bell state).
The first-order derivative of $V$ yields
\begin{eqnarray}\label{3}
\dot{V}&=&-2\sum_{k}f_k(t)\textrm{\textbf{Im}}[\langle
\psi|\psi_T\rangle \langle \psi_T|H_k|\psi\rangle]    \nonumber\\
   &=&-2\sum_{k}|\langle  \psi|\psi_T\rangle|f_k(t)\textrm{\textbf{Im}}
   [e^{i\arg\langle\psi|\psi_T\rangle}
\langle \psi_T|H_k|\psi\rangle],   \nonumber\\
\end{eqnarray}
where $\textrm{\textbf{Im}}(...)$ stands for the imaginary part of
$(...)$ and $\textrm{arg}\langle\psi|\psi_T\rangle$ is the phase
difference between $|\psi\rangle$ and $|\psi_T\rangle$.
Thus, the condition $\dot{V}\leq0$ can be satisfied naturally if we
choose the control fields
\begin{eqnarray}  \label{5}
f_k(t)=-B_{k}\textrm{\textbf{Im}}[e^{i\arg\langle\psi|\psi_T\rangle}
\langle \psi_T|H_k|\psi\rangle],  ~~ k=1,2,
\end{eqnarray}
where the constant $B_k>0$. Fig. \ref{fig:03}(a)-(b) demonstrates how the system
arrives at the Bell state $|\psi_T\rangle$ by control fields $f_k(t)$
when the initial state is $|0001\rangle$.
In particular we have designed the control
fields for both QDs to be the same,
i.e., $f_2(t)=f_1(t)$.  We further observe that the total manipulation
time is related to the constant $B_1$, e.g., Fig. \ref{fig:03}(c)-(d) are plotted
with $B_1=300$. An inspection of Fig.
\ref{fig:03}(a)-(d) shows that the amplitude of control
fields is time-dependent, which may make difficult for experimental
realization. Actually, by the virtue of the fundamental principle of Lyapunov
control, the dynamics performance is insensitive to the amplitude of
control fields. Instead, it depends sharply on the sign of
control fields. Due to this flexibility feature, the time-dependent amplitude
of control fields can be replaced by square pulses as follows,
\begin{eqnarray}\label{8}
f_k(t)=\left \{
\begin{array}{rl}
    F,~~~ f_k(t)>0, \\
    -F,~~~f_k(t)<0. \\
\end{array}
\right.
\end{eqnarray}
Fig. \ref{fig:03} (e)-(h) show the performance to realize the Bell
state $|\psi_T\rangle$ by square pluses, where the form of control fields are much simpler than that
given  by Eq. (\ref{5}). Furthermore, the control time
is also shortened by square pluses.

\begin{figure}[h]
\centering
\includegraphics[scale=0.4]{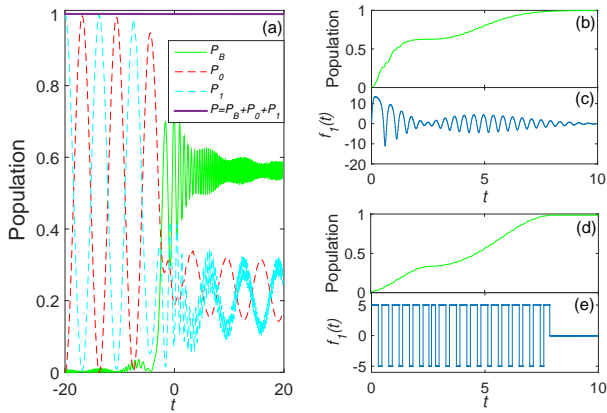}
\caption{Population of the Bell state $|\psi_T\rangle$ (green-solid
line) as a function of  time in presence of the Josephson coupling
$E_J=0.5$. Panel (a) demonstrates the results by adiabatic passage,
while panels (b)-(e) demonstrate the system evolution by Lyapunov
control. The red-dash (cyan-dash) line denotes the
population of $|0000\rangle$ ($|0001\rangle$). $P\simeq1$
(purple-solid) shows it has two major distinct physical mechanisms
during the evolution,  since the population of other states almost
vanish. The other parameters are the same as in Fig. \ref{fig:03}.
(a) $T=20$. The control field in panel (c) is given by Eq.(\ref{5})
with $B_1=300$,  while it is calculated by Eq.(\ref{8}) with $F=5$
in panel (e). } \label{fig:04}
\end{figure}

Now we turn to the case when existing the Josephson coupling.
The Hilbert space is now spanned by $\{|0000\rangle,
|0001\rangle, |0110\rangle, |1010\rangle, |1100\rangle,
|1101\rangle\}$, as sketched in Fig. \ref{fig:01}(b).  We have
neglected the higher-energy excited states once more. At first we
decrease the chemical potentials of two QDs adiabatically when the
initial state is $|0001\rangle$. The dynamical behaviors are plotted
in Fig. \ref{fig:04}(a). It can be observed that adiabatic passage
is invalid for perfectly generating the Bell state $|\psi_T\rangle$
since the population cannot reach one. The reason can be
found as follows. The evolution process mainly contains two physical
mechanisms: (i) the Rabi oscillation between $|0000\rangle$ and
$|0001\rangle$ for the existence of Josephson coupling; (ii) the
population adiabatic transfer from $|0001\rangle$ to the Bell state
$|\psi_T\rangle$. The population adiabatic transfer dominates only
in the middle-stage and the throughout existence of Rabi oscillation
lead to failure generation of the Bell state $|\psi_T\rangle$. But
when employing Lyapunov control, we find in Fig. \ref{fig:04}(b)-(e)
that whether system exists the Josephson coupling makes no
difference in the entanglement generation since the Bell state
$|\psi_T\rangle$ is still the system eigenstate. The only difference
from the aforementioned case is the shape of control field $f_1(t)$.

\begin{figure}[h]
\centering
\includegraphics[scale=0.45]{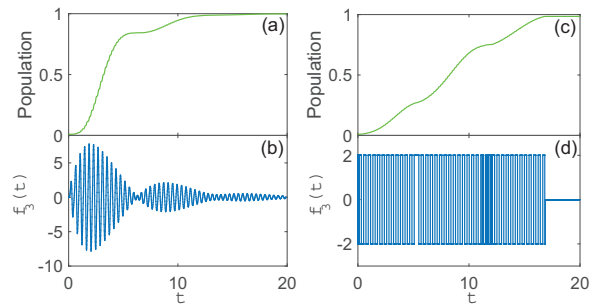}
\caption{The population of the Bell state $|\psi_T\rangle$ as a function  of
time, where the control Hamiltonian is the
Cooper pair exchange. (b) $B_3=100$. (d) $F=2$. }  \label{fig:05}
\end{figure}

Alternatively, we can utilize the Cooper pair exchange between the
TS wire and the bulk superconductor to be the control Hamiltonian in
Lyapunov control, i.e.,
\begin{eqnarray}   \label{8a}
H_3=\cos\hat{\phi}=\frac{1}{2}(e^{-i\hat{\phi}}+e^{i\hat{\phi}}).
\end{eqnarray}
As shown in Fig. \ref{fig:05}, the control field $f_3(t)$ designed by
Eq.(\ref{5}) or Eq.(\ref{8}) can steer the system into the Bell
state $|\psi_T\rangle$.
Thus we can generate the Bell state $|\psi_T\rangle$ by modulating the
strength of Cooper pairs exchange between the TS wire and the bulk
superconductor.

\section{Crossed Andreev reflection scheme}

Crossed Andreev reflection (CAR), also known as nonlocal Andreev
reflection, occurs when two spatially separated  electrodes in
normal state form two separate junctions with a superconductor.
In our model a CAR process refers to the situation that
two electrons  from different QDs  tunnel  into the TS wire to form
a Cooper pair \cite{plugge15,recher01,lesovik01,nilsson08}. Inversely, two
electrons separated from a Cooper pair in the TS wire would tunnel
into the  QDs separately. Note that local Andreev reflection is absent since
it cannot occupy two electrons for the same QD in the Coulomb block regime.

Recently, it has been shown that a  strong and long-range
coupling between two spatially separated QDs can be induced by a
superconductor via CAR \cite{leijnse13}.  Here in presence of MFs and charging
energy $E_c$, we demonstrate that CAR can also be used to prepare
long-range entanglement in two QDs when the parameters are $n_g=0$ and
$\epsilon_1=-\epsilon_2=\epsilon$.
The Hilbert space spanned by low-energy eigenstates is
$\{|0000\rangle, |0001\rangle, |011-1\rangle, |0110\rangle,
|1010\rangle, |101-1\rangle, |110-1\rangle,|1100\rangle\}$. Here we
use the same notation $|n_1n_2n_fN_c\rangle$ as in   the first case.
With this basis,  the Hamiltonian $H_0$ can be written as a
$8\times8$ matrix. As illustrated  in Fig. \ref{fig:01}(c), the
entanglement can not be prepared without the Josephon coupling
$E_J$. In other words,  it requires Cooper pair exchange between the
TS wire and the bulk superconductor. Hence we refer this process as
crossed Andreev reflection scheme. Although the eigenstates can be
calculated analytically, the expressions are tedious. Hence  we
adopt numerical solutions to discuss the occupation of the  two
lowest eigenstates of the Hamiltonian $H_0$.

\begin{figure}[h]
\centering
\includegraphics[scale=0.4]{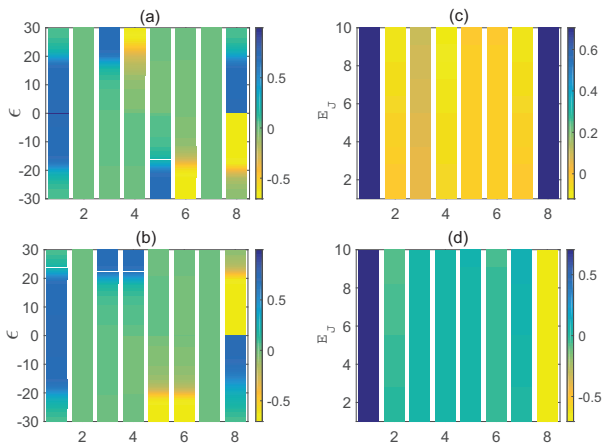}
\caption{The probability amplitude of each basis appear in the two
lowest eigenstates as a function of the chemical potential (panels
(a)-(b)) and the Josephson coupling constant(panels (c)-(d)).  The
horizontal axis denotes the eight basis ordered as \{$|0000\rangle$,
$|0001\rangle$, $|011-1\rangle$, $|0110\rangle$, $|1010\rangle$,
$|101-1\rangle$, $|110-1\rangle$, $|1100\rangle$\}.}  \label{fig:06}
\end{figure}

\begin{figure}[h]
\centering
\includegraphics[scale=0.4]{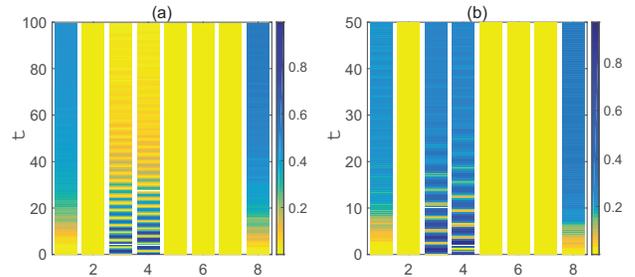}
\caption{The population on each basis as a function of time. The
eight basis are ordered as \{$|0000\rangle$, $|0001\rangle$,
$|011-1\rangle$, $|0110\rangle$, $|1010\rangle$, $|101-1\rangle$,
$|110-1\rangle$, $|1100\rangle$\} in horizontal axis. The initial
state is $|0110\rangle$ and the target state is
$|\psi_T^{+}\rangle=\frac{1}{\sqrt{2}}(|00\rangle+|11\rangle)|00\rangle$.
The control Hamiltonians are (a) $H_1=d_1^{\dag}d_1$,
$H_2=d_2^{\dag}d_2$,  (b) $H_3=\cos\hat{\phi}$. All control fields
are designed by Eq.(\ref{5}). }  \label{fig:07}
\end{figure}

Fig. \ref{fig:06} shows that the eigenstate is nearly the
Bell state
$|\psi_T^{\pm}\rangle=(|00\rangle\pm|11\rangle)|00\rangle$ when
$|\epsilon|<20$,  and there is not an eigenstate that approximately
equals one of the basis. This indicates that  it is difficult to
prepare  the Bell state by adiabatic passage. To prepare  the Bell
state $|\psi_T^{+}\rangle$ by Lyapunov control, we first explore a
situation that the control Hamiltonians are the particle number of
the two QDs (i.e., $H_n=d_n^{\dag}d_n$, $n=1,2$). With the control
fields given by Eq.(\ref{5}), we plot the results  in Fig.
\ref{fig:07}(a). We find  that the final state can be approximately
expressed as
$|\psi\rangle\approx\frac{1}{\sqrt{2}}(|00\rangle+|11\rangle)|00\rangle$,
which is actually the Bell state. For the situation that the control
Hamiltonian is the Cooper pair exchange (i.e.,
$H_3=\cos\hat{\phi}$), the results are plotted in Fig.
\ref{fig:07}(b). We find that the final state is not a Bell state,
indicating it is fail to prepare the Bell state $|\psi_T^{+}\rangle$
in this situation. Further observations reveal that the populations
on the basis \{$|0000\rangle$, $|011-1\rangle$, $|0110\rangle$,
$|1100\rangle$\} nearly equal each other, which implies  that
 the Bell state $|\psi_T^{+}\rangle$ may be obtained
by  measuring the parity of Dirac fermion. Indeed, further
examination yields that if $n_f=0$, the final state collapses to
$|\psi_T^{+}\rangle=\frac{1}{\sqrt{2}}(|00\rangle+|11\rangle)|00\rangle$.
If $n_f=1$, the final state collapses to
$|\psi\rangle\approx\frac{1}{\sqrt{2}}(|011-1\rangle+|0110\rangle)$,
leading to  the failure for generation of the Bell state $|\psi_T^{+}\rangle$.

\section{Scheme with intra-dot spin flip}

In the last  section, we propose a scheme to entangle  two QDs
regardless of electron spin states. We can also employ the spin up
and spin down to encode quantum information due to their  long
coherence time \cite{loss98,petta05,koppens05}. In the following, we
will investigate this issue.

\begin{figure}[h]
\centering
\includegraphics[scale=0.45]{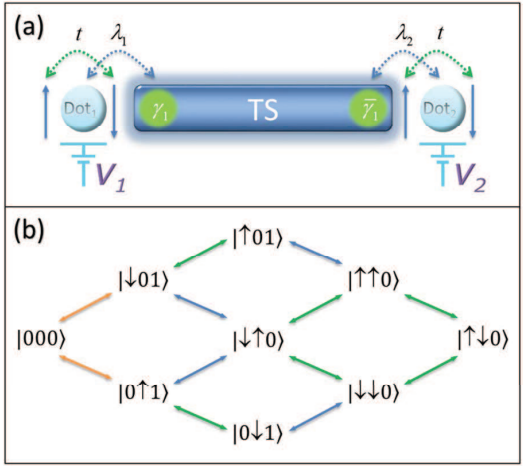}
\caption{ The schematic setup consisting of a TS  wire tunnel
coupling to two QDs. Only spin down (up) electron in the first
(second) QD can be tunnel coupled to the TS wire since the
left (right) MF $\gamma_1$ ($\bar{\gamma}_1$) is spin down (up) as
well. At the same time, there exists the intra-dot spin flip
process in two QDs.  (b) The transition between distinct states.
The green line represents the intra-dot spin flip process.}
\label{fig:01b}
\end{figure}

As sketched in Fig. \ref{fig:01b}(a), the system differs not so much
from the setup in Fig. \ref{fig:01}(a). It does not have the bulk
superconductor and the gate charge. In particular, in order
to keeps spin degeneracy in QDs, we do not apply magnetic fields but
employ a magnetic insulator contacting with the TS wire to induce
the effective Zeeman coupling \cite{fu08,sau10}. The Hamiltonian
 reads \cite{flensberg11,leijnse11,tewari08,ke15}
\begin{eqnarray}   \label{9}
H_{0}'&=&\sum_{n=1}^{2}\left(\sum_{\nu}\epsilon d_{n\nu}^{\dag}d_{n\nu}+U_{n}
d_{n\uparrow}^{\dag}d_{n\uparrow}
d_{n\downarrow}^{\dag}d_{n\downarrow}
+td_{n\uparrow}^{\dag}d_{n\downarrow}\right)        \nonumber\\
&&+\lambda_{1}(f^{\dag}+f)d_{1\downarrow}
+\lambda_{2}(f^{\dag}-f)d_{2\uparrow}+h.c.
\end{eqnarray}
The first term describes the energy of two QDs with the chemical
potential $\epsilon$ for spin projection $\nu=\uparrow,\downarrow$.
The second term describes two electrons occupying the same QD with
  Coulomb interaction $U_n$. In the Coulomb block regime, the
electron can only occupy single fermion level in the two QDs, and we
focus on this regime in the following. The third term
describes the intra-dot spin flip   with   strength $t$, which stems
from spin-orbit interactions and has been studied  in Ref.
\cite{khaetskii00}. Note that the spin of QD is no more a good
quantum number as the transitions are allowed between distinct
spin states. Since the spin flip term plays an essential role in
the entanglement generation, as shown by the green line  in Fig.
\ref{fig:01b}(b), we will call this process  the intra-dot spin flip
scheme. The fourth (fifth) term describes the tunnel coupling
between the Dirac fermion   and the spin down (up) electron in the
first (second) QD with strength $\lambda_1$ ($\lambda_2$).

In general, the MFs always have a definite  spin
polarization at the two ends of TS wire. Thus the wire can only
send or receive electrons with the same spin polarization. The spin
polarization of MFs are determined by the boundary between
topological and non-topological regions
\cite{sticlet12,kjaergaard12}, which are  antiparallel at the two
ends of TS wire \cite{oreg10,fu08} in an ideal case. We should
emphasize that the spin polarization  is not antiparallel in  real
cases, which would cause an error in the entanglement preparation
\cite{leijnse12a}. Nevertheless, we can manipulate  the chemical
potentials around the ends of TS wire to achieve nearly  perfect
antiparallel spin polarization. This is the reason why in the fourth
(fifth) term we have assumed the electron state is spin down (up) on
the left (right) MF $\gamma_1$ ($\bar{\gamma}_1$), and it can
be only coupled to the spin down (up) electron in the first (second)
QD via tunneling.

When we consider the even-parity case, the Hilbert
space is spanned by \{$|000\rangle$, $|\downarrow01\rangle$,
$|0\uparrow1\rangle$, $|\uparrow01\rangle$, $|0\downarrow1\rangle$,
$|\downarrow\uparrow0\rangle$, $|\uparrow\uparrow0\rangle$,
$|\uparrow\downarrow0\rangle$, $|\downarrow\downarrow0\rangle$\}. We
have employed the label $|n_1n_2n_f\rangle$ to describe system
state, where $n_{1}(n_2)=0,\uparrow,\downarrow$ denotes the electron
state in the first (second) QD and $n_f$ denotes the number of Dirac
fermion combined by MFs. Note that the Hamiltonian $H_0'$ can be
represented by a $9\times9$ matrix in these basis. As the analytical expressions of
system eigenstates are involved, we choose to show the behaviors
of two lowest eigenstates by numerical calculations.

\begin{figure}[h]
\centering
\includegraphics[scale=0.4]{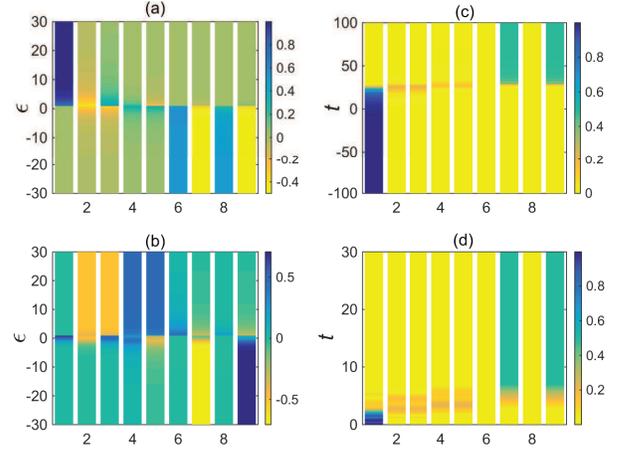}
\caption{The amplitude of each basis in two lowest eigenstates as a
function of the chemical potential in panels (a)-(b). The horizontal
axis denotes the nine basis ordered as \{$|000\rangle$,
$|\downarrow01\rangle$, $|0\uparrow1\rangle$, $|\uparrow01\rangle$,
$|0\downarrow1\rangle$, $|\downarrow\uparrow0\rangle$,
$|\uparrow\uparrow0\rangle$, $|\uparrow\downarrow0\rangle$,
$|\downarrow\downarrow0\rangle$\}. One can find that the amplitude
mainly locates at basis $|000\rangle$ when $\epsilon\gg0$ in panel (a), while the
amplitude mainly equal-weighted locates at basis
$|\uparrow\uparrow0\rangle$ and $|\downarrow\downarrow0\rangle$ when
$\epsilon\ll0$ in panel (b). The population of each basis as a function of
time in panels (c)-(d). $\epsilon=-10$. The control fields
are (c) $f(t)=-\frac{2}{5}t+20$ in adiabatic passage, (d) designed
by Eq.(\ref{5}) with $B_1=1000$ in Lyapunov control. After
completing controls, the amplitude of final state mainly
equal-weighted locates at basis $|\uparrow\uparrow0\rangle$ and
$|\downarrow\downarrow0\rangle$.}  \label{fig:09}
\end{figure}

Fig. \ref{fig:09}(a)-(b) describes the amplitude of each basis in
two lowest eigenstates as a function of the chemical potential. It
demonstrates that the  eigenstate approximately equals to the state
$|000\rangle$ when the chemical potential $\epsilon\gg0$, while the
eigenstate nearly becomes the Bell state
$|\psi_T'\rangle=(|\uparrow\uparrow\rangle-|\downarrow\downarrow\rangle)|0\rangle$
when the chemical potential $\epsilon\ll0$. In addition to this we
find in Fig. \ref{fig:01b}(b) that the transition path
$|000\rangle\leftrightarrow|\uparrow\uparrow0\rangle$ and the
transition path
$|000\rangle\leftrightarrow|\downarrow\downarrow0\rangle$ are
completely symmetric so that it is possible to generate the Bell
state by adiabatic passage. The main control procedures are as
follows. The system is initialized in the state $|000\rangle$ with
large chemical potential. Then we adiabatically decrease the
chemical potential. The system would evolve to the Bell state
$|\psi_T'\rangle=(|\uparrow\uparrow\rangle-|\downarrow\downarrow\rangle)|0\rangle$
finally, which is plotted in Fig. \ref{fig:09}(c). An universal
drawback of adiabatic passage is that it takes long control time in
order to meet adiabatic condition. To reduce the control time we
turn to Lyapunov control, where the control fields are designed by
Eq.(\ref{5}) instead of decreasing gradually (in adiabatic control).
The dynamics evolution is demonstrated in Fig. \ref{fig:09}(d). One
can find the total control time of implementing the Bell state
$|\psi_T'\rangle$ by Lyapunov control is much shorter than that by
adiabatic passage.

\section{Scheme beyond intra-dot spin flip}

\begin{figure}[h]
\centering
\includegraphics[scale=0.45]{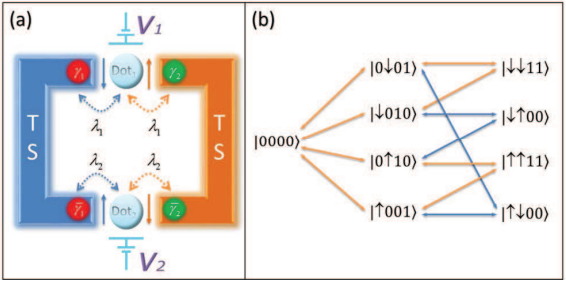}
\caption{(a) The schematic setup consisting of two  TS
wires tunnel coupling to two QDs. (b) The transition between
distinct states.}  \label{fig:01c}
\end{figure}

We can  also prepare  the Bell states based on a scheme beyond
intra-dot spin flip. The system considered is composed of two TS
wires coupling to two common QDs, as illustrated  in Fig.
\ref{fig:01c}(a). The Hamiltonian describing such a system  reads
\cite{flensberg11,leijnse11,tewari08}
\begin{eqnarray}    \label{10}
H_{0}''&=&\sum_{n=1}^{2}\Big{(}\sum_{\nu}\epsilon d_{n\nu}^{\dag}d_{n\nu}
+U_{n}d_{n\uparrow}^{\dag}d_{n\uparrow}
d_{n\downarrow}^{\dag}d_{n\downarrow}+\lambda_{n+1}[f_{n}^{\dag}-   \nonumber\\
&&(-1)^{n-1}f_{n}]d_{n+1\uparrow}
+\lambda_{n}[f_{n}^{\dag}+(-1)^{n-1}f_{n}]d_{n\downarrow}+h.c.\Big{)},  \nonumber\\
\end{eqnarray}
where the notation is the same as in Eq.(\ref{9}), and we adopt the
convention $n \mod 2$ if $n>2$. We use
$|n_1n_2n_{f_1}n_{f_2}\rangle$ to denote system state, where
$n_{1}(n_2)=0,\uparrow,\downarrow$ stands for an electron state in
the first (second) QD and $n_{f_1}=f_1^{\dag}f_1$
($n_{f_2}=f_2^{\dag}f_2$) denotes the number of Dirac fermion
defined by MFs in the left (right) TS wire.

\begin{figure}[h]
\centering
\includegraphics[scale=0.4]{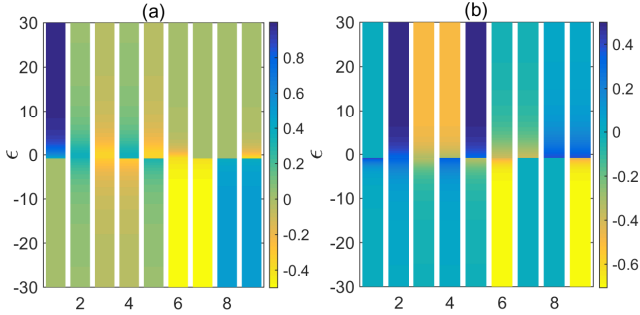}
\caption{The amplitude of each basis in two lowest
eigenstates as a function of the chemical potential.  The horizontal
axis denotes the nine basis ordered as \{$|0000\rangle$,
$|0\downarrow01\rangle$, $|\downarrow010\rangle$,
$|0\uparrow10\rangle$, $|\uparrow001\rangle$,
$|\downarrow\downarrow11\rangle$, $|\uparrow\downarrow00\rangle$,
$|\downarrow\uparrow00\rangle$, $|\uparrow\uparrow11\rangle$\}.}
\label{fig:12}
\end{figure}

\begin{figure}[h]
\centering
\includegraphics[scale=0.4]{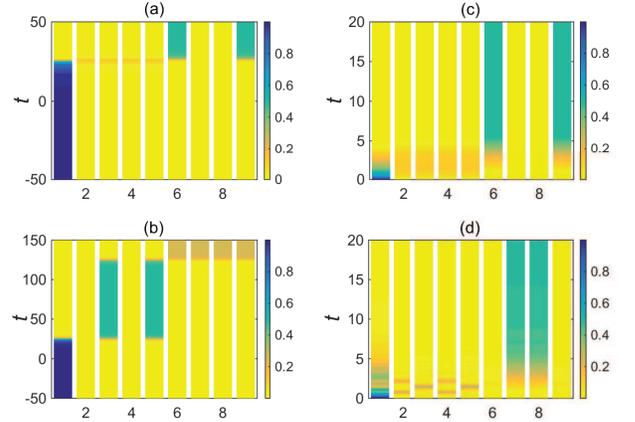}
\caption{The population on each basis as a  function of time when
the initial state is $|0000\rangle$.  The horizontal axis denotes
the nine bases  ordered as \{$|0000\rangle$,
$|0\downarrow01\rangle$, $|\downarrow010\rangle$,
$|0\uparrow10\rangle$, $|\uparrow001\rangle$,
$|\downarrow\downarrow11\rangle$, $|\uparrow\downarrow00\rangle$,
$|\downarrow\uparrow00\rangle$, $|\uparrow\uparrow11\rangle$\}. It
presents two distinct adiabatic methods: (a) simultaneously
decreasing the chemical potential of two QDs with the same rate,
i.e., $f_1(t)=f_2(t)=-\frac{6}{5}t+30$. (b) decreasing the chemical
potential of two QDs in turns, i.e., $f_1(t)=-\frac{6}{5}t+30,
t\in[-50,50]$, and $f_2(t)=-\frac{6}{5}(t-100)+30, t\in[50,150]$.
$f_1(t)=f_2(t)=0$ for other evolution time. Deterministic generation
of two Bell states $|\psi_1\rangle$ and $|\psi_2\rangle$ in Lyapunov
control while the target state is (c)
$|\psi_1\rangle=(|\downarrow\downarrow\rangle+|\uparrow\uparrow\rangle)|11\rangle$,
(d)
$|\psi_2\rangle=(|\uparrow\downarrow\rangle+|\downarrow\uparrow\rangle)|00\rangle$.}
\label{fig:13}
\end{figure}

Similar to the analysis presented in the last section, we first
analyze the behaviors of the two lowest eigenstates as a function of
chemical potential with numerical simulations. When $\epsilon\gg0$,
the  eigenstate is very close to the basis $|0000\rangle$ [see Fig.
\ref{fig:12}(a)]. When $\epsilon\ll0$, the   eigenstate  is
approximately  a superposition of $|\downarrow\downarrow11\rangle$
and $|\uparrow\uparrow11\rangle$ with almost equal weight [see Fig.
\ref{fig:12}(b)]. It is believed that the system can be steered into
the Bell state by adiabatic passage when the initial state is
$|0000\rangle$. Here we explore this issue by two different
adiabatic passages. The first   is to simultaneously decrease the
chemical potentials of the two QDs with the same rate, as shown in
Fig. \ref{fig:13}(a). The system finally is driven into the Bell
state
$|\psi_1\rangle=\frac{1}{\sqrt{2}}(|\downarrow\downarrow\rangle+|\uparrow\uparrow\rangle)|11\rangle$.
Another method is to adiabatically decrease the chemical potential
of the first QD while the chemical potential of the second QD
remains unchanged. After completing this operation, the system state
becomes
$|\psi'\rangle=\frac{1}{\sqrt{2}}(|\downarrow010\rangle+|\uparrow001\rangle)$.
We then adiabatically decrease the chemical potential of the second
QD while the chemical potential of the first QD remains unchanged.
The results are shown in Fig. \ref{fig:13}(b). The final state of
system becomes
$|\psi\rangle=\frac{1}{2}(|\uparrow\downarrow\rangle+|\downarrow\uparrow\rangle)|00\rangle-
\frac{1}{2}(|\downarrow\downarrow\rangle+|\uparrow\uparrow\rangle)|11\rangle$,
which is actually not the Bell state of the two QDs. However, with
the help of measurement results on the Dirac fermion parity (e.g., $n_{f_1}$),
the system would  collapse into   Bell state
$|\psi_2\rangle=\frac{1}{\sqrt{2}}(|\uparrow\downarrow\rangle+|\downarrow\uparrow\rangle)|00\rangle$
if $n_{f_1}=0$, while the system would  collapse to   Bell state
$|\psi_1\rangle=\frac{1}{\sqrt{2}}(|\downarrow\downarrow\rangle+|\uparrow\uparrow\rangle)|11\rangle$
if $n_{f_1}=1$. Remarkably, the control results are quite different
for the two adiabatic passages. This originates from the fact that
it exists quantum destructive interference in the first adiabatic passage. To
clarify this point, we examine the transition paths between
$|0000\rangle$ and $|\uparrow\downarrow00\rangle$ in Fig.
\ref{fig:01c}(b), i.e.,
\begin{eqnarray}
|0000\rangle&\xlongleftrightarrow{-\lambda_2}&|0\downarrow01
\rangle \xlongleftrightarrow{\lambda_1}|\uparrow\downarrow01\rangle, \nonumber\\
|0000\rangle&\xlongleftrightarrow{\lambda_1}&|\uparrow001\rangle
\xlongleftrightarrow{\lambda_2}|\uparrow\downarrow01\rangle. \nonumber
\end{eqnarray}
Due to the presence of minus sign in the coupling constant, these
two transition paths destructively  interfere  in the first method
(The mechanism  is same for the transition paths between
$|0000\rangle$ and $|\downarrow\uparrow00\rangle$). But this effect
cannot happen in the second method.

From the aspect of Lyapunov control, since both $|\psi_1\rangle$ and
$|\psi_2\rangle$ are the  eigenstates of the system when
$\epsilon\ll0$ ($|\psi_2\rangle$ is not shown in Fig. \ref{fig:12}),
it is then possible to generate   Bell states by Lyapunov control,
where the control fields are designed by Eq.(\ref{5}). The results
are presented  in Fig. \ref{fig:13}(c)-(d). Compared with the
adiabatic control,  a  specific type of Bell states (e.g.,
$|\psi_1\rangle$ and $|\psi_2\rangle$) can be prepared without
measurement.

\section{Discussion and conclusion}

Before concluding, we discuss  the validity of the
assumptions made in our model  and the experimental feasibility of
our proposal.

The first assumption in this work is that there are no other
quasiparticles in the TS wire except for MFs. This assumption is
true  if the gap $\Delta$ in the superconductor  is sufficiently
large. Recent experiments
\cite{das2012,mourik12,deng12,finck13,higginbotham15} report the
observation of MFs in the TS wire (e.g., InAs or InSb nanowire) and
the superconducting  gap $\Delta$ is in the order of 0.1-1 meV,
which is much larger than thermal fluctuations $\sim k_BT$ with
temperature $\lesssim$100 mK.

In the second assumption, we suppose that both QDs are in
the Coulomb block regime. Experimentally, the two QDs can be
constructed at the two ends of the same TS wire.  The electrostatic
gates  underneath the TS wire create a confinement potential for
electron to form a  QD. It have been demonstrated that the Coulomb
interaction $U$ in such QD device is in the order of 1-10 meV
\cite{fasth07,nilsson09}, which is much larger than the
superconducting  gap $\Delta$ and the thermal fluctuation at
operating temperature. Therefore the QDs in the Coulomb block regime
is realistic  in our model.

Besides, the charging energy $E_c$ is also an essential
parameter for generating long-range entanglement, which is in the
order of  $1$ meV \cite{higginbotham15}. To meet the condition
$E_c<\Delta$, one can increase the length of TS wire or capacitance
to reduce the charging energy $E_c$. If the charging energy $E_c$ is
larger than the superconducting  gap $\Delta$,  quasiparticles would
appear in the TS wire, which participate in the entanglement
preparation. When such quasiparticles are in the bulk of TS wire,
i.e., no remarkable  overlap with MFs, they do not have effect  on
the Bell state preparation. Nevertheless, when the quasiparticles
are located near the ends of TS wire, it may be invalid to prepare
the Bell states by adiabatic passage or Lyapunov control.

Since the chemical potential of QDs can be modulated by
electrostatic gates, we mainly discuss  how to change the gate
voltage to simulate the control fields. To prepare the Bell states
by adiabatic passage, we just need to decrease the gate voltage with
time linearly, which is easily realized  in experiments. In Lyapunov
control, it may not be easy to manipulate the gate voltage
to simulate the time-dependent amplitude of control fields
[e.g., Fig. \ref{fig:03}(b)], but it is quite easy to realize
the square pulses required by the control fields
[e.g., Fig. \ref{fig:03}(f)]. One may notice that
the tunnel coupling $\lambda$ is fixed in our calculation. In fact,
the tunnel coupling depends on both the chemical potential of QDs
and tunneling barriers. As the tunnel coupling changes with the
chemical potentials of QDs, but one can employ additional electrostatic gates on
tunnel barriers to keep the tunnel coupling fixed \cite{perge10}.

Finally, we  discuss the effect   of decoherence on the
preparation of Bell states   in charge qubit, since the spin qubit
has a coherence time (in the order of microseconds) longer than
charge qubit. Due to the electron-phonon interactions, an intrinsic
decoherence mechanism in QDs, the lifetime of charge qubit is in the
order of $16$ ns \cite{petta04,petersson10}. In realistic
situations, the tunnel coupling $\lambda$ can be modulated by
electrostatic gates and reach   the order of 1-10 $\mu$eV, so the
total manipulation time is in the order of 6-60 ns by adiabatic
passage. Hence  the adiabatic passage would be  invalid for
Bell states preparation if the tunnel coupling $\lambda$ is too weak.
However the manipulation time is in the order of 0.7-7 ns by Lyapunov
control, which is much shorter  than the lifetime of charge qubit.
Therefore   Lyapunov control is feasible for entangled state
preparation when the tunnel coupling is not very large.

In conclusion, by the teleportation scheme, the crossed Andreev
reflection scheme, the intra-dot spin flip scheme, and the scheme
beyond intra-dot spin flip, we show how to entangle two QDs
mediated by a pair of MFs. The Bell states can be prepared in both the
charge degrees and spin degrees of QDs by Lyapunov control.  In
contrast, we compare our results with those by adiabatic passage.
The Lyapunov control manifests advantages over adiabatic passage at
flexibility designing control fields and accelerating control time.

In the teleportation scheme, the system can be driven into Bell
state $|\psi_T\rangle=\frac{1}{\sqrt{2}}
({|10\rangle-|01\rangle})|10\rangle$ by both adiabatic passage
and Lyapunov control when the initial state is $|0001\rangle$.
When the Josephson coupling is taken into account, it is not available
to prepare Bell states perfectly by adiabatic passage due to the
existence of Rabi oscillation. However, the Lyapunov
control can still work well by modulating the shape of control
fields. In addition, we find that the Cooper pair exchange can also be
served as the control Hamiltonian to generate Bell states. In the
crossed Andreev reflection scheme, since it does not exist a low-energy
eigenstate whose amplitude mainly locates at one of the basis,
we directly turn to Lyapunov control. The results show that the
system can reach the Bell state
$|\psi_T^{+}\rangle=\frac{1}{\sqrt{2}}(|00\rangle+|11\rangle)|00\rangle$
when we regulate the chemical potentials of two QDs.
However, if we choose the Cooper pair exchange as control
Hamiltonian, whether it can generate the Bell state successfully
or not depends on parity measurement results of the Dirac fermion formed by MFs.

As to the entanglement in the spin degree of freedom, we have
studied the system in the presence of intra-dot spin flip process.
Through exploring the low-energy eigenstates of the system, it
demonstrates that one can achieve the Bell state
$|\psi_T'\rangle=(|\uparrow\uparrow\rangle-|\downarrow\downarrow\rangle)|0\rangle$
by both adiabatic passage and Lyapunov control. For the scheme
beyond intra-dot spin flip, the system consists of two TS wires and
two QDs.  Interestingly, we find that the results are quite
different in distinct adiabatic passages, i.e., decreasing the
chemical potentials of two QDs simultaneously or in turns. This
diversity originates from whether existing quantum destructive
interference in adiabatic evolution. Finally we have showed that the
system can be certainly driven into distinct Bell states by Lyapunov
control.

\section*{ACKNOWLEDGEMENT}

We thank H. F. L\"u for helpful discussions. This work is
supported by the National Natural Science Foundation
of China (Grant Nos. 11175032 and 61475033).

\end{document}